\documentclass[article]{nature}

\usepackage{graphicx}
\usepackage{color}

\title{Tunable Multifunctional Topological Insulators in Ternary Heusler Compounds}

\author{Stanislav Chadov$^{1}$, Xiaoliang Qi$^{2,4}$,  J\"urgen
  K\"ubler$^3$, Gerhard~H.~Fecher$^1$, Claudia Felser$^1$ \& Shou Cheng Zhang$^4$}

\begin{document}

\maketitle

\begin{affiliations}
 \item Institut f\"ur Anorganische Chemie und Analytische Chemie,
             Johannes Gutenberg - Universtit\"at,  55099 Mainz,
             Germany
\item Microsoft Research, Station Q, Elings
Hall, University of California, Santa Barbara, CA 93106, USA
\item Institut f\"{u}r Festk\"{o}rperphysik, Technische Universit\"{a}t Darmstadt, 64289 Darmstadt, Germany
\item Department of Physics, McCullough Building, Stanford University, Stanford, California 94305-4045, USA
\end{affiliations}

\begin{abstract}

Recently  the   Quantum  Spin  Hall  effect  (QSH)   was  theoretically predicted and
experimentally realized  in a quantum  wells based on binary semiconductor
HgTe.\cite{Bernevig06,Koenig07,Dai08} QSH state and topological insulators
are the new states of quantum matter interesting both for fundamental condensed
matter physics and material
science.\cite{Qi2010,Kane05,Bernevig06a,Bernevig06,Koenig07,Fu07,Hsieh08,Dai08,Zhang09,Xia09,Chen09}
Many of  Heusler compounds  with C1$_b$
structure   are  ternary  semiconductors   which  are   structurally  and
electronically related to the  binary semiconductors.  
The  diversity of Heusler materials
opens  wide possibilities  for  tuning  the band  gap  and setting  the
desired   band  inversion   by  choosing   compounds   with  appropriate
hybridization  strength  (by lattice  parameter)  and  the magnitude  of
spin-orbit coupling (by the atomic charge).
Based on  the first-principle  calculations we demonstrate  that
around fifty Heusler compounds  show the band inversion similar  to
HgTe. The topological state in these zero-gap semiconductors can be created
by applying strain or by designing an appropriate quantum well structure,
similar to the case of HgTe. Many of  these ternary  zero-gap semiconductors  (LnAuPb,
LnPdBi,  LnPtSb and LnPtBi) contain the rare earth element Ln which
can realize  additional  properties
ranging from  superconductivity (e.~g. LaPtBi\cite{Goll08})
to magnetism (e.~g.     GdPtBi\cite{Canfield91})
and heavy-fermion    behavior (e.~g. YbPtBi\cite{Fisk91}).
These properties can open new research directions in realizing the quantized
anomalous Hall effect and topological superconductors.

\end{abstract}

According to their electronic structure all bulk materials are divided
into metals, which have a finite electron density at the Fermi
energy and insulators, which show the band gap. Recently a new class
of the so-called {\em topological} states has emerged -  the Quantum
Spin Hall (QSH) state in two dimensions and its generalization in
three dimensions. The corresponding materials, called topological
insulators (TI) have been predicted theoretically and recently
studied
experimentally.\cite{Bernevig06,Koenig07,Fu07,Hsieh08,Zhang09,Xia09,Chen09}
The TI is a new state of quantum matter with a full insulating gap
in the bulk, but with topologically protected gapless surface or
edge states  on the boundary.  Research on  TI's has  attracted
considerable attention  due to  both  fundamental   interest to a
new state  of  matter as well as   its  possible application in new
spintronic devices.\cite{Qi2010} All the TI's discovered  so far
were either   alloys   (Bi$_{1-x}$Sb$_x$\cite{Fu07,Hsieh08})  or
binary compounds (HgTe,\cite{Bernevig06,Koenig07} Bi$_2$Se$_3$,
Sb$_2$Te$_3$ and Bi$_2$Te$_3$.\cite{Zhang09,Xia09,Chen09})  This
letter reports about tunable {\em multifunctional}   TI's  within
the  class   of  ternary semiconducting Heusler compounds.  The
great diversity of these materials (more than
 200 semiconductors among 500 Heusler compounds) opens wide possibilities for tuning
their bandstructure comparing to binary compounds and enables 
an effective search of the optimal TI material
for applications.

In particular, several compounds,  YPtSb, YPdBi and ScAuPb are found
at their experimental lattice constants~\cite{VC91} 
close to the border between the trivial and topological insulator,
with all relevant bands degenerated at the $\Gamma$ point (see Fig.~3(a)).
Such a material can be easily transformed from a trivial to a topological
insulator and vice versa by a small variation of the lattice
constant (by applying pressure or growing the film on the
appropriate substrate).  Many TI candidate compounds  (LnAuPb, LnPdBi, LnPtSb and
LnPtBi) containing the rare earth elements Ln with strongly
correlated $f$-electrons can exhibit various conventional orders,
such as magnetism,\cite{Canfield91} superconductivity or heavy
fermion behavior.\cite{Fisk91} Such conventional orders in TI's enable the
realization of many novel topological effects and exotic particles,
such as the image monopole effect,\cite{Qi09} axion\cite{li2010} and
Majorana fermion.\cite{Fu08} Thus the ternary TI's
are multifunctional and can be exploited to design new devices. 
Combinations of the ternary trivial/topological insulators (such as
ScPtSb/ScPtBi similar to CdTe/HgTe) can be used as quantum well 
devices for the QSH.

Ternary Heusler compounds of  X$_2$YZ or XYZ composition (with X, Y -
the transition or rare earth metals and Z - the main group element)
form the class of materials which are well-suited for
various spintronic applications.\cite{Felser07}
The semiconducting nature of these compounds arise due to the strong
tendency towards covalent bonding.  From basic structural and bonding 
considerations the X$_2$YZ Heusler
compounds (L2$_1$ structure)  with 18 or 24 valence
electrons\cite{Galanakis02}  and XYZ Heuslers (C1$_b$) with 
18 valence electrons\cite{Jung00,Galanakis02a,Kandpal06}
are expected to exhibit a gap at the Fermi energy.
In the following we will focus on the XYZ Heuslers (also called half-Heuslers).
The relation to the classical semiconductors is easy to understand
for such materials as e.\,g., ${\rm LiZnAs}$. More
surprisingly, LaPtBi (sometimes also named LaBiPt)
is a semiconductor formed by  three metallic elements!
Typically the XYZ half-Heuslers can be viewed as comprised from
X$^{n+}$ ion ``stuffing'' the zincblende YZ$^{n-}$ sublattice (here Li$^{+}$
stuffed in [ZnAs]$^{-}$ or La$^{3+}$ stuffed in [PtBi]$^{3-}$
sublattice) where the number of valence electrons  associated with
YZ$^{n-}$ is equal to 18 ($d^{10}+s^2+p^6$). 18-electron compounds
are closed shell species, non-magnetic and semiconducting.

Figure~1 illustrates a comparison of the zincblende  and the
Heusler structures. The additional stuffed rare earth
atoms are shown as orange spheres. Similar to the binary
semiconductors, the band gap can be tuned by the  electronegativity
difference of the constituents and the lattice
constant\cite{Kandpal06} in a wide range from about 4~eV (LiMgN)
down to zero (LaPtBi). The semiconducting Fe$_2$VAl\cite{Kato01} and
ZrNiSn\cite{Sakurada05} Heusler compounds  in addition demonstrate
the excellent thermoelectric properties, just like the known
TI's Bi$_2$Te$_3$ and Bi$_2$Se$_3$.\cite{Middendo73,Hor09}
To understand ternary compounds as TI's, we compare them to the binary
 compounds studied in the literature. For example, CdTe is
a trivial semiconductor, whereas HgTe is topological
system.\cite{Bernevig06,Koenig07,Fu07,Dai08} Such drastic change in
properties occurs due to interplay of the spin-orbit coupling
(produced by the heavy Hg and Te atoms) and the degree of hybridization
(controlled by the lattice constant).
The same considerations apply to all ternary semiconducting Heusler compounds. Further
we verify this based on first-principle
bandstructure calculations by using the
fully-relativistic version\cite{PYA}  of the standard
LMTO (Linearized Muffin-Tin Orbitals) approach. The exchange-correlation 
part of the effective potential was treated using Vosko-Wilk-Nusair
parameterization\cite{VWN80} of the  Local Density Approximation (LDA).  
The following example (Figure~2)  compares  calculated bandstructures
 of CdTe and HgTe  with  ScPtSb  and ScPtBi.
For direct comparison with model calculations\cite{Bernevig06}
we mark the relevant bands carrying $\Gamma_6$ representation as
blue and $\Gamma_8$ as red.  As it follows, the bandstructures of
these ternary compounds reveal  clear fingerprints: both CdTe and
ScPtSb exhibit a direct gap at the $\Gamma$ point between the
conduction  (blue)  and the valence (red) bands of $\Gamma_6$ and
$\Gamma_8$ symmetries, respectively.
On the other hand, the bandstructures of  HgTe and  ScPtBi
exhibit the same band inversion:   $\Gamma_6$ (blue)  is now situated below
$\Gamma_8$ (red) which remains at the Fermi energy. 
This is the necessary condition for the TI state, since it changes the parity 
 of the wave function compared to CdTe or ScPtSb 
(for details see the work of Dai and coworkers\cite{Dai08}).   

To illustrate the impressive number of new topological compounds and the possibility
to tune the $\Gamma_6$-$\Gamma_8$ band ordering in this class of materials,
 we have performed similar calculations for all relevant Heuslers containing
Sc, Y, La, Lu and Th. The energy difference between $\Gamma_6$
and $\Gamma_8$ bands  as a function of lattice constant is
shown on Figure~3(a). Each subgroup  (e.g. Ln~=~Sc, Y, La, Lu) is marked by a
certain color. Compounds containing
 $f$-electrons, such as Pr Tb, etc. (except of Ce) are  not presented here,
since  their partially filled  $f$-electron shells with strong
correlations require  special treatment that goes beyond standard
LSDA approach. The compounds with ${E_{\Gamma_6}-E_{\Gamma_8}>0}$
 are  trivial insulators, whereas those with
 ${E_{\Gamma_6}-E_{\Gamma_8}<0}$ are the TI candidates. The latter group
 consists of  the  zero-gap semiconductors with double-degenerated $\Gamma_8$
point at the Fermi energy.  It follows that all existing Heuslers with zero
band gap at the Fermi energy (around 50 compounds by including the
rare earth based) under certain conditions
will demonstrate the same type of band inversion
as HgTe. Indeed, the increase of the lattice constant reduces the
hybridization and closes the non-zero band gap.
Combined with sufficiently strong spin-orbit coupling it leads to a
pronounced $\Gamma_6$-$\Gamma_8$ band inversion, which is the key
to realize the TI state. Figure~3(b) demonstrates the
$\Gamma_6$-$\Gamma_8$ difference as a function of the average spin-orbit
coupling expressed by the sum of the nuclear charges of the
elements in the unit cell, i.e. Z(X)+Z(Y)+Z(Z). The latter 
appears to be a suitable order parameter which sorts the materials almost 
along the straight line.   As it follows from Figure~3, the
combinations  of Pt with Bi in LnPtBi or Au with Pb in LnAuPb
series always lead to the inverted band structure.
There is an additional advantage of Heusler materials
which also follows from  Figure~3(a): due to the large amount of
compounds with different gap values it is easy to construct a
quantum well comprised of the trivial and  topological parts
with  well-matching lattice constants, similar to the HgTe/CdTe quantum
well. The appropriate pairs can
be chosen from the candidates situated in the  middle area of
Figure~3(a) along the same vertical line,  since  the transition from
trivial to a topological behavior as a function of lattice constant
appears to be rather smooth in average. The relevant combinations, 
for example, are ScPdBi/ScPtBi, YPdSb/YPtSb and LuAuSn/LuPtBi.

Although the analogy with HgTe is quite clear and convincing, we
would like to demonstrate the topological nontrivial nature of the
zero gap Heuslers in a more rigorous way. On this purpose, we
study the topological phase transition between trivial and inverted
phases in YPdBi, one the materials which sits
right at the border between the trivial and topological insulator
states (see Fig.~3).  Figure.~4(a) demonstrates the
corresponding transition in details for YPdBi induced by
a small variation of the lattice constant. The insets show the corresponding
bandstructures.  At the equilibrium lattice constant
 a linear dispersing Dirac cone is formed by the
 $\Gamma_8$ light-hole and $\Gamma_6$ bands,  degenerate
with the $\Gamma_8$ quadratic heavy-hole band. Small variation of the lattice
constant (by about ${\pm 0.3~\%}$) substantially influences the band
structure near the Fermi energy: the compression leads to a trivial
state with the band gap ${E_{\Gamma_6}-E_{\Gamma_8}\approx 0.07}$~eV, whereas the expansion leads
to a zero-gap inverted state with a difference ${E_{\Gamma_6}-E_{\Gamma_8}\approx -0.07}$~eV.
Analogous effect is achieved by scaling the spin-orbit coupling magnitude $\lambda$ 
at the critical lattice constant (Fig.~4(b)). The
suppression of spin-orbit coupling ($\lambda<1$) drives the system into a
trivial state while the increase $(\lambda>1)$ leads to a topological insulator.

Furthermore, for HgTe it has been proposed that a uniaxial strain
can lift the degeneracy between light-hole and heavy-hole subbands
of $\Gamma_8$, which drives this zero-gap semiconductor into a real
TI phase.\cite{Fu07,Dai08} Similar effects
can be studied by our {\em ab-initio} calculations in Heusler
compounds.  Figure~5 shows the YPdBi bandstructure calculated for a
small tetragonal strain applied along [001]-direction  
reducing the $c/a$ ratio by 3~\%.
One can see that both normal (Fig.~5(a)) and inverted (Fig.~5(c))
systems are gapped. The gapless Dirac cone at the ``critical'' lattice
constant (Fig.~5(b)) remains robust, while the quadratic heavy-hole band is
pushed away from Fermi level. It should be noticed that the
dispersion at the critical point is linear along all directions in
the momentum space around $\Gamma$, although the linear
region is very small along the directions perpendicular to the
strain direction. Consequently, such a transition driven by the change of lattice
constant corresponds to the sign change of the mass term in 3D Dirac
equation, which is exactly the theory of the topological phase
transition between trivial and topological insulators (this is a
similar approach as that used for 2D HgTe quantum
well\cite{Bernevig06}). Thus we conclude that the inverted
half-Heuslers are topologically nontrivial.

The proposed materials can be tuned from a trivial to
a topological insulator mainly in two different ways: (i)~by 
variation of the lattice constant (applying pressure or growing the
material on appropriate substrate); (ii)~by substitution of elements
(varying their electronegativities or the strength of spin-orbit
coupling). (iii)~The devices allow for the additional  options to
manipulate the electronic structure, such as to switch the
borderline compound from trivial to topological by applying a gate
voltage or  by constructing the quantum well structures.

After the initial discovery of the QSH systems, the current research
is now focused on the proximity effect between the TI's 
 and other forms of ordered state, such as the magnetism
and the superconductivity.\cite{Qi2010} However, the presently known
TI's only become magnetic or superconducting when
doped with extrinsic elements such as Mn, Fe and Cu. In contrast,
the TI's based on Heusler compounds naturally
include the $f$-shell rare earth elements intrinsically, together
forming a stoichiometric system. Besides the chemical functions
(passing of the three electrons to the zincblende lattice and
determining the lattice size) the additional open $f$-shell element
renders  multifunctionality by providing the coexistence of
conventional ordering with TI state, that is necessary for the
realization of novel topological effects and the new extended
applications. Here we list several examples of such multifunctional
materials. (i)~Bulk magnetism found in LnPtBi (Ln~=~Nd, Sm
Gd, Tb, Dy)\cite{Canfield91} may realize the dynamical
axion,\cite{li2010} which is the spin-wave excitation with a
topological coupling with electromagnetic field. Such an effect
provides a new design of tunable optical modulator. (ii)~The
heavy-fermion behavior in YbPtBi\cite{Fisk91} may realize the
recently proposed topological Kondo insulator.\cite{Dzero09} 
(iii)~The superconductivity in
the noncentrosymmetric low-carrier LaPtBi system.\cite{Goll08} The
absence of inversion symmetry is theoretically proposed to support
topological superconductivity.\cite{Qi09B} Thus it is interesting to
investigate whether the superconductivity in LaPtBi is topological.

\begin{addendum}
  \item This work is supported by ARO, grant number W911NF-09-1-0508.
   Financial support by the Deutsche Forschungsgemeinschaft DFG
(Research unit FOR~559, project P~07) is gratefully acknowledged.
  \item[Correspondence] Correspondence and requests for materials
 should be addressed to Claudia Felser (felser@uni-mainz.de) and Shou
 Cheng Zhang (sczhang@stanford.edu).
 \end{addendum}

\begin{figure}
\centering
\begin{minipage}[t]{1.0\textwidth}
\includegraphics[width=0.97\textwidth]{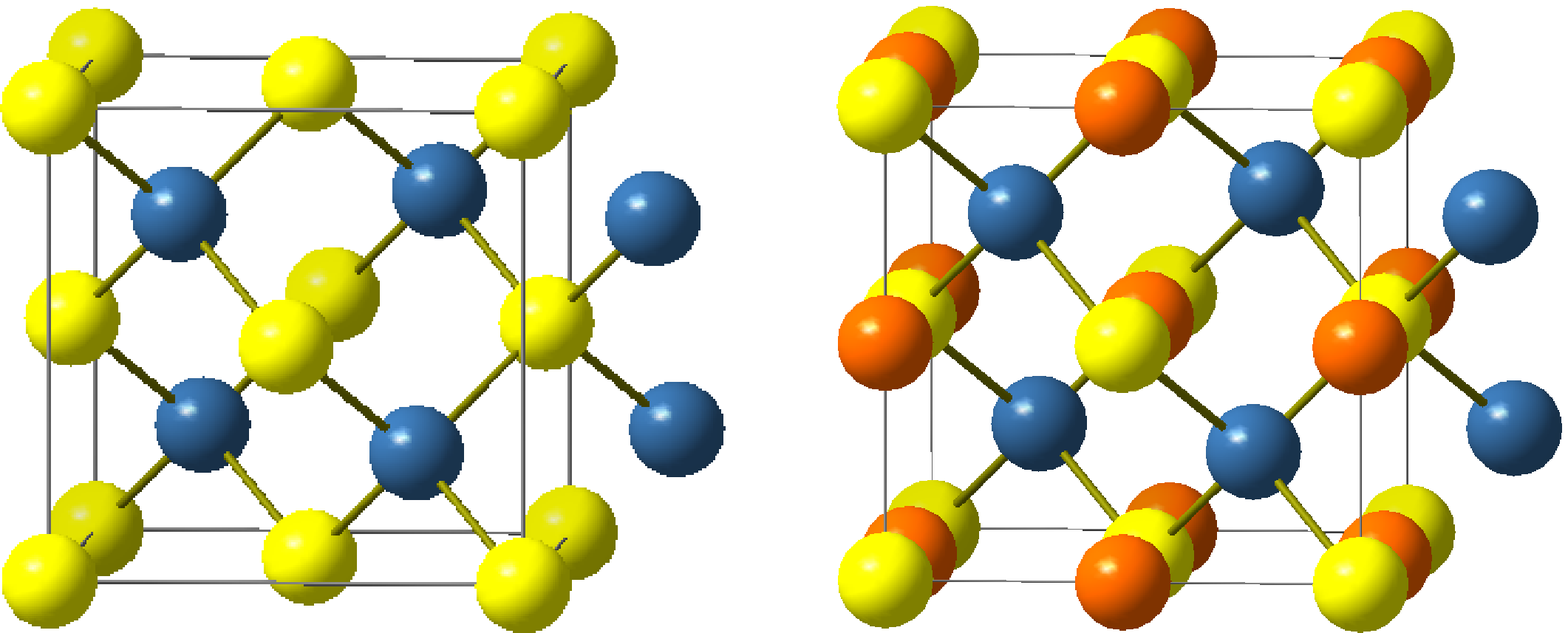}
\end{minipage}
\caption{Zincblende (YZ) structure (left) in comparison to
   the C1$_b$ structure (right) of Heusler (XYZ) compounds. Yellow and
   blue spheres correspond to the main group (Z) and transition (Y) elements,
   respectively. The orange spheres in C1$_b$ stand for the additional
   stuffing (X) element.}
\end{figure}

\begin{figure}
\centering
\begin{minipage}[c]{1.0\textwidth}
\includegraphics[width=1.0\textwidth]{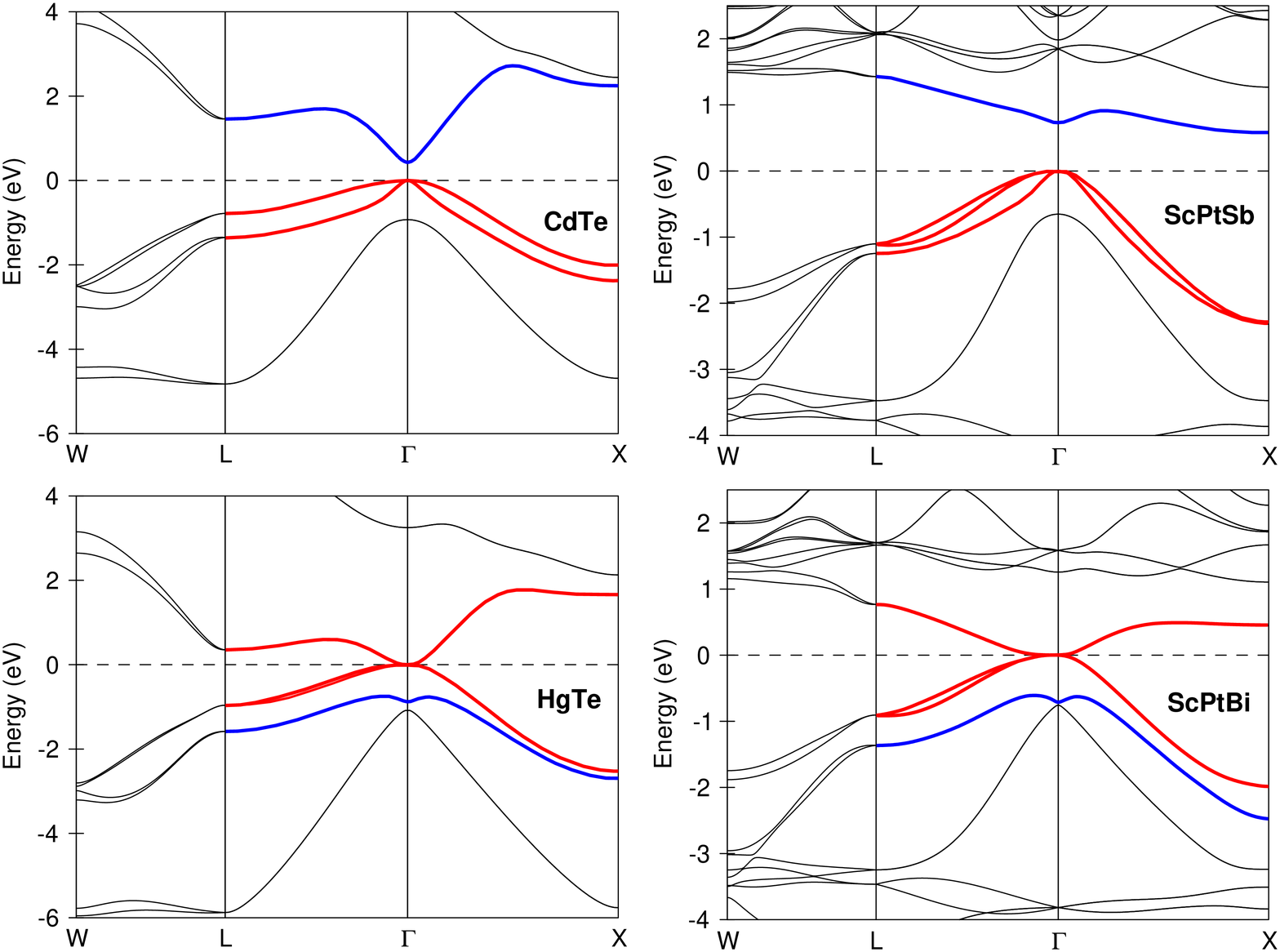}
\end{minipage}
\caption{Bandstructures of CdTe and HgTe compared to ScPtSb and
    ScPtBi Heuslers. Red color mark the bands with $\Gamma_8$
    symmetry, blue - with $\Gamma_6$. Comparison reveals obvious
    similarity between binary systems and their ternary equivalents: 
    both CdTe and ScPtSb are trivial
    semiconductors with $\Gamma_6$ situated above $\Gamma_8$ which sits
    at the Fermi energy (set to zero). Both HgTe and ScPtBi are
    topological with inverted bands order, the band with $\Gamma_6$
    symmetry is situated below $\Gamma_8$.}
\end{figure}

\begin{figure}
\centering
\begin{minipage}[t]{1.0\textwidth}
\includegraphics[width=1.0\textwidth]{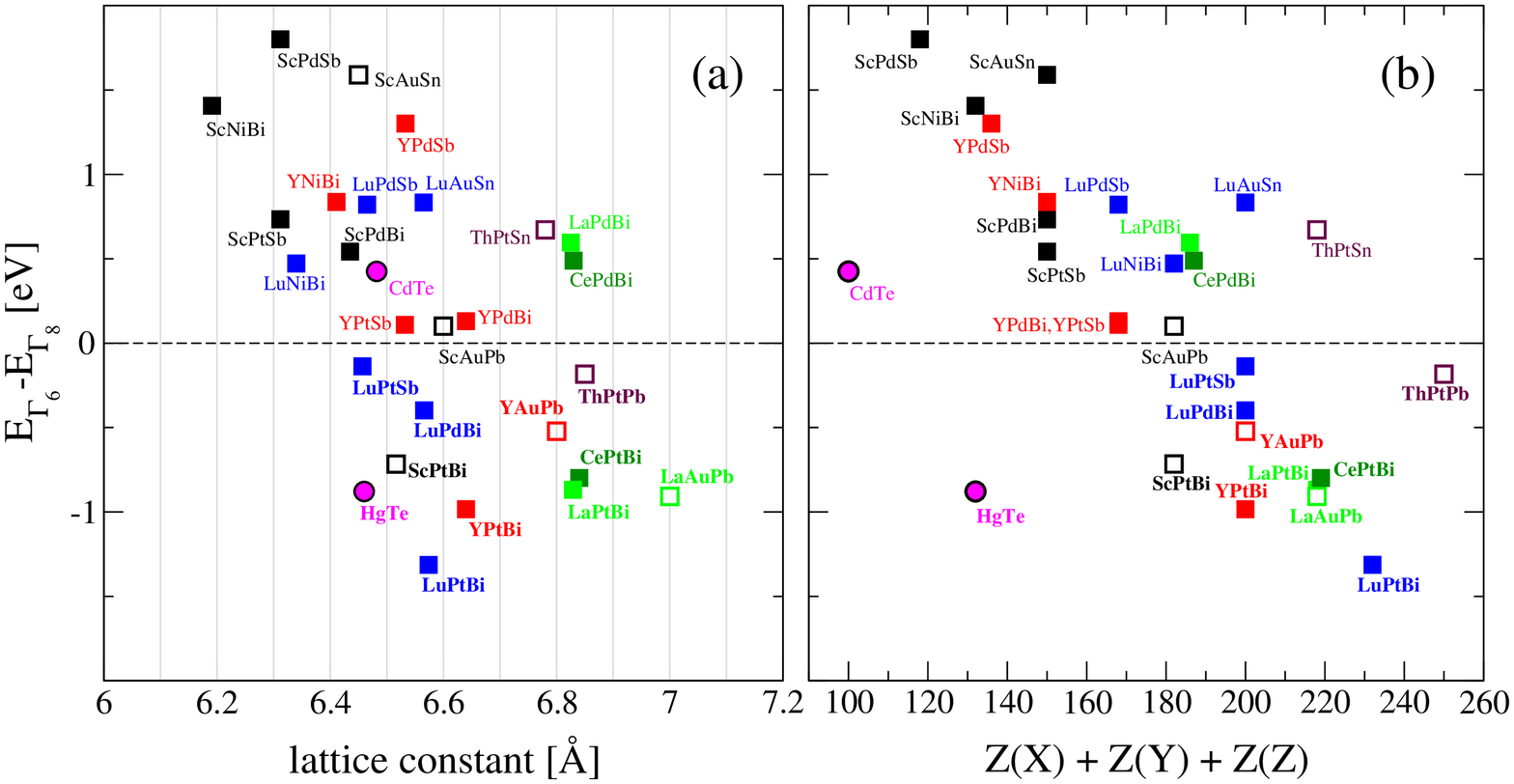}
\end{minipage}
\caption{  $E_{\Gamma_6}-E_{\Gamma_8}$ difference  
calculated for various Heuslers
 at their experimental lattice constants. HgTe and CdTe
  binaries are shown for comparison. Hollow squares mark the systems not
  reported in the literature. (a)
  $E_{\Gamma_6}-E_{\Gamma_8}$ difference
 as a function of
the lattice constant. Pairs of materials with  well-matching
lattices for the QSH quantum wells can be easily picked up along the
same vertical lines. The borderline compounds (between trivial and
topological) insulators (YPtSb, YPdBi, ScAuPb) are situated 
closer to the zero horizontal line.  
(b) $E_{\Gamma_6}-E_{\Gamma_8}$ difference as a function of the average spin-orbit coupling
strength represented by a sum of the nuclear charges over the
unit cell.}
\end{figure}

\begin{figure}
\centering
\begin{minipage}[t]{1.0\textwidth}
\includegraphics[width=1.0\textwidth]{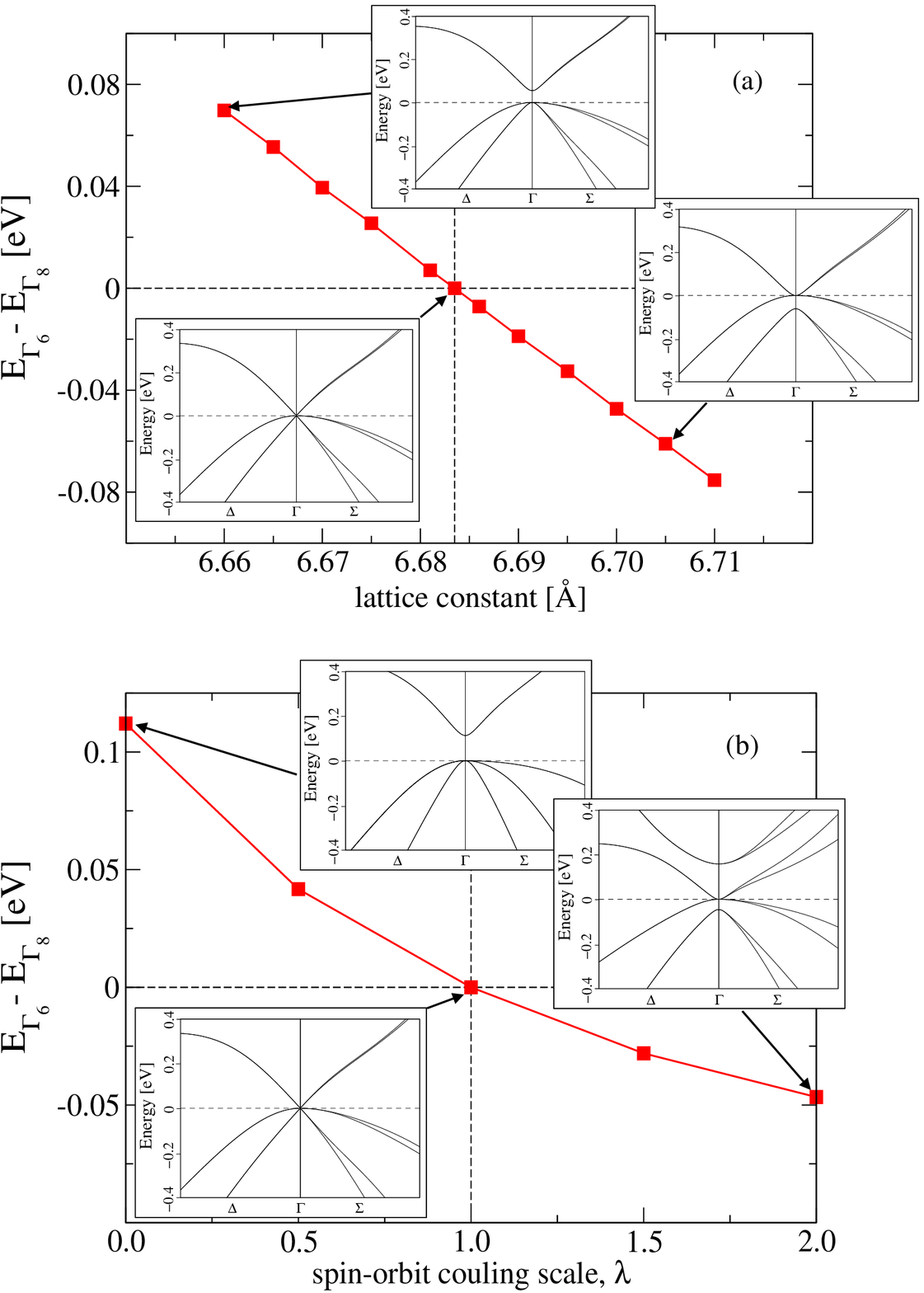}
\end{minipage}
\caption{$E_{\Gamma_6}-E_{\Gamma_8}$ difference for YPdBi.
The insets show the bandstructures
     at the marked points along $\Delta$ and $\Sigma$ symmetry 
     directions of the Brillouine zone.
(a) $E_{\Gamma_6}-E_{\Gamma_8}$  changes by scaling the   
    lattice constant from positive (trivial insulator) 
    to negative (topological insulator). The borderline (critical) 
    lattice constant (marked by dashed vertical line) corresponds to the
     zero gap with a Dirac cone at the Fermi energy. (b)
     Analogously, by scaling the spin-orbit coupling at the critical
     lattice constant, the system
     can be transformed  from trivial ($\lambda<1$) to topological
     insulator ($\lambda>1$).}
\end{figure}

\begin{figure}
\centering
\begin{minipage}[t]{1.0\textwidth}
\includegraphics[width=1.0\textwidth]{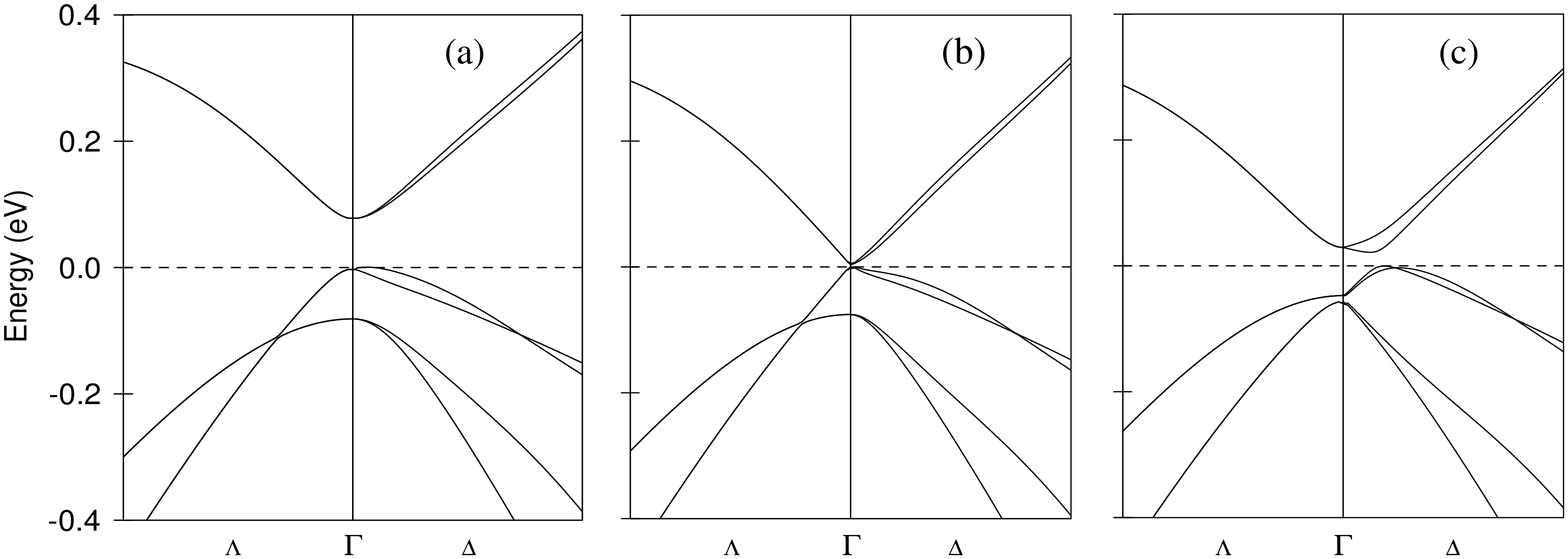}
\end{minipage}
\caption{Bandstructure of YPdBi under tetragonal strain ($c/a=0.97$).
  (b) At the critical lattice constant the strain shifts down the
  heavy-hole band with parabolic dispersion, leaving the single Dirac
  cone at the Fermi energy. This effect is especially pronounced along
  $\Lambda$ direction, parallel to the strain. We note, that along
  $\Delta$ (perpendicular to the strain) the bands at the Fermi energy
  still have a linear dispersion near the $\Gamma$ point, although on
  much smaller scale.  Variation of the lattice constant (by $\pm$0.4~\%)
  around the critical value leads to a trivial state by compression
  (a) or topological by expansion (c).}
\end{figure}

\end{document}